\documentclass[12pt]{iopart}

\usepackage{iopams}
\usepackage{amssymb}
\usepackage{textcomp}
\usepackage{graphicx}
\usepackage[T1]{fontenc}

\begin{document}
%=======================================================================================

\title{Trapping atoms in the evanescent field of laser written wave guides}

\author{D Juki\'c$^1$, A Moqanaki$^2$, P Walther$^2$, A Szameit$^3$, T Pohl$^1$ and J B G\"otte$^{1,4}$}
\address{$^1$Max Planck Institute for the Physics of Complex Systems, N\"othnitzer Str. 38, 01187 Dresden, Germany}
\address{$^2$Faculty of Physics, University of Vienna, Boltzmanngasse 5, A-1090 Vienna, Austria}
\address{$^3$Institute of Applied Physics, Abbe Center of Photonics, Friedrich-Schiller-University Jena, Max-Wien-Platz 1, 07743 Jena, Germany}
\address{$^4$School of Physics and Astronomy, University of Glasgow, University Avenue, Glasgow G12 8QQ, UK}

\date{\today}

\begin{abstract}
We analyze evanescent fields of laser written waveguides and show that they can be used to trap atoms close to the surface of an integrated optical atom chip.
In contrast to subwavelength nanofibres it is generally not possible to create a stable trapping potential using only the fundamental modes.
This is why we create a stable trapping potential by using two different laser colors, such that the waveguide supports two modes for the blue detuned laser, while for the red detuned light the waveguide has only a single mode.
In particular, we study such a two-color trap for Cesium atoms, and calculate both the potential and losses for the set of parameters that are within experimental reach.
We also optimize system parameters in order to minimize trap losses due to photon scattering and tunneling to the surface.
\end{abstract}

%\pacs{05.30.-d, 03.75.Kk, 67.85.De}
\maketitle
%\narrowtex%\newpage
%=======================================================================================

%%%
%%% Introduction
%%%

\section{Introduction}
\label{sec:intro}

The attractions of integrated atom-light interfaces lie in the control and enhancement of optical fields offered by photonics combined with the versatility of atomic systems.
With the help of such interfaces a range of new possibilities in atomic and optical physics can be explored, which include a pathway to implement scalable quantum networks \cite{Kimble:Nature:2008} and the ability to engineer light-matter interactions in quantum many-body physics \cite{Goban+:NatComm5:2014, Pichler+:PRA91:2015}.
The photonic part of the interface is commonly used to create a strong optical confinement, as is achieved for sub-wavelength fibres \cite{Balykin+:PRA70:2004,Vetsch+:PRL104:2010} or photonic crystals \cite{Goban+:NatComm5:2014}.
While this approach allows for the generation of strong coupling between light and atoms, it also requires a dedicated and often expensive production process which stands against the requirements of a cost-effective and scalable interface and begs the question whether there migth not be a simpler approach.

This is why we investigate the possibility to create a scalable light matter interface using the evanescent field of wave guides which have been written into a dielectric medium by means of femtosecond laser pulses.
This process is fast and cost effective and allows for the creation a variety of geometries \cite{SzameitNolte:JPB43:2010}.
Laser written waveguides have been used to study and experimentally verify numerous physical phenomena with classical light which often have analogy with coherent quantum effects found in atomic or condensed matter physics \cite{Lederer+:PhysRep463:2008, Longhi:LPR3:2009, SzameitNolte:JPB43:2010}.
Arrays of these waveguides are usually written into glass such that the optical modes of neighboring waveguides are coupled via bulk evanescent fields, thereby realizing tight-binding models with tunable hoppings \cite{SzameitNolte:JPB43:2010}.
Laser written waveguides also present an ideal platform to explore optical nonlinearities, leading for example, to the observation of various two-dimensional soliton solutions \cite{Szameit:PRL:2007, Lederer+:PhysRep463:2008}.
In addition to this, recent experiments include demonstration of pseudomagnetic fields and photonic Landau levels \cite{Rechtsman+:NatPhot7:2012}, photonic Floquet topological insulators \cite{Rechtsman+:Nature496:2013}, and generation of high-order photonic entangled W-states \cite{Graefe+:NatPhot8:2014}.

In most experiments with laser written waveguides, the evanescent field leaking outside of the bulk medium has not been exploited.
One exception is an optofluidic sensor in microchannels etched inside silica glass structure \cite{Maselli+:OE17:2009}.
On the other hand, proposals and experiments to use evanescent fields in order to achieve coherent light-atom coupling are already well established in other photonic systems.
In particular, evanescent fields have been used for laser trapping and optically interfacing atoms around dielectric nanofibers \cite{Balykin+:PRA70:2004, Vetsch+:PRL104:2010, LeKien+:PRA70:2004, Lacroute+:NJP14:2012, Goban+:PRL109:2012}
More recently, strong single atom and photon interactions have been achieved in evanescent field of microtoroidal resonators \cite{Alton+:NatPhys7:2010}, and in the near field of nano-photonic crystals \cite{ Goban+:NatComm5:2014, Tiecke+:Nature508:2014}.

\begin{figure}
 	\includegraphics[width=0.4\textwidth]{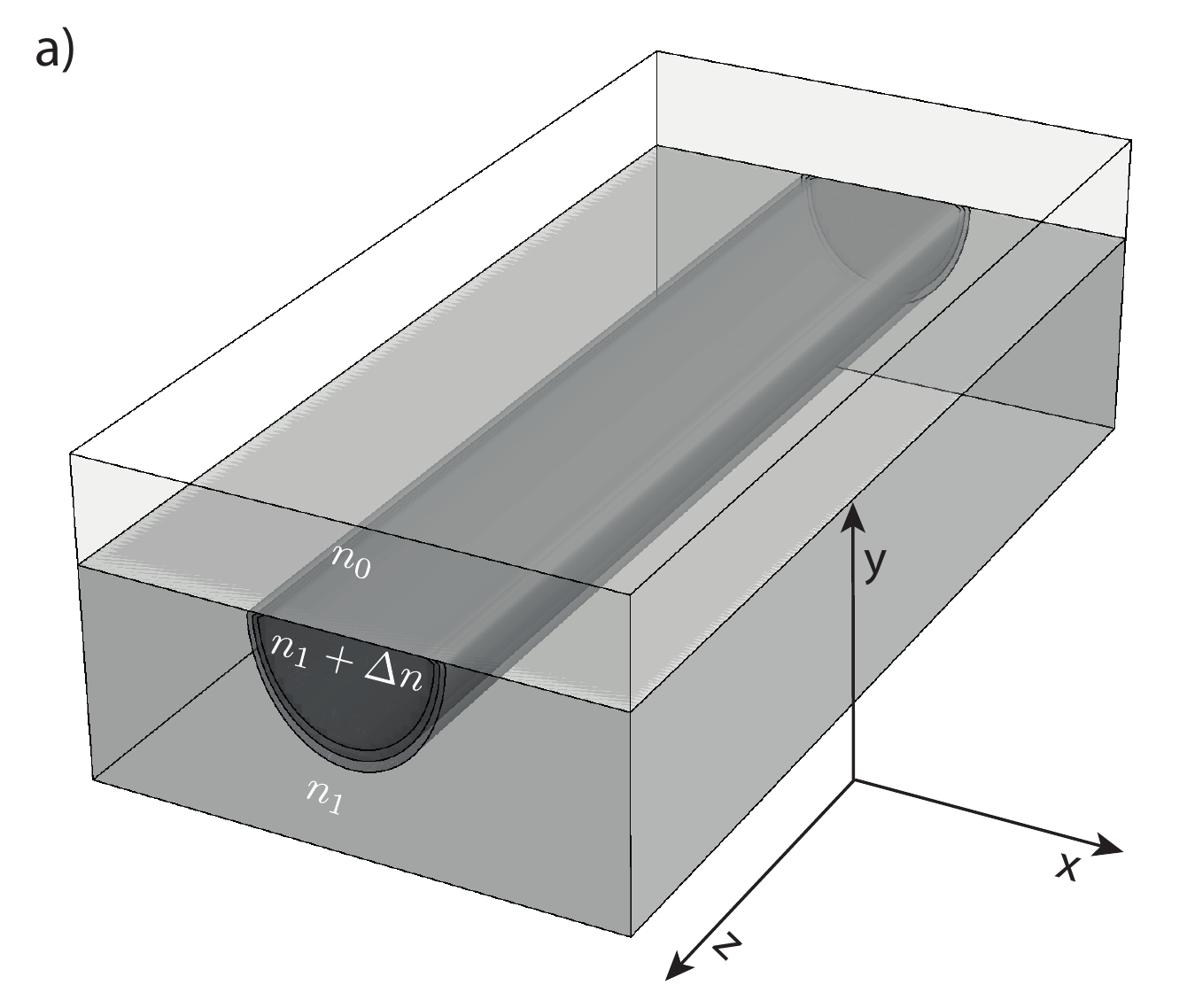}
 	\includegraphics[width=0.45\textwidth]{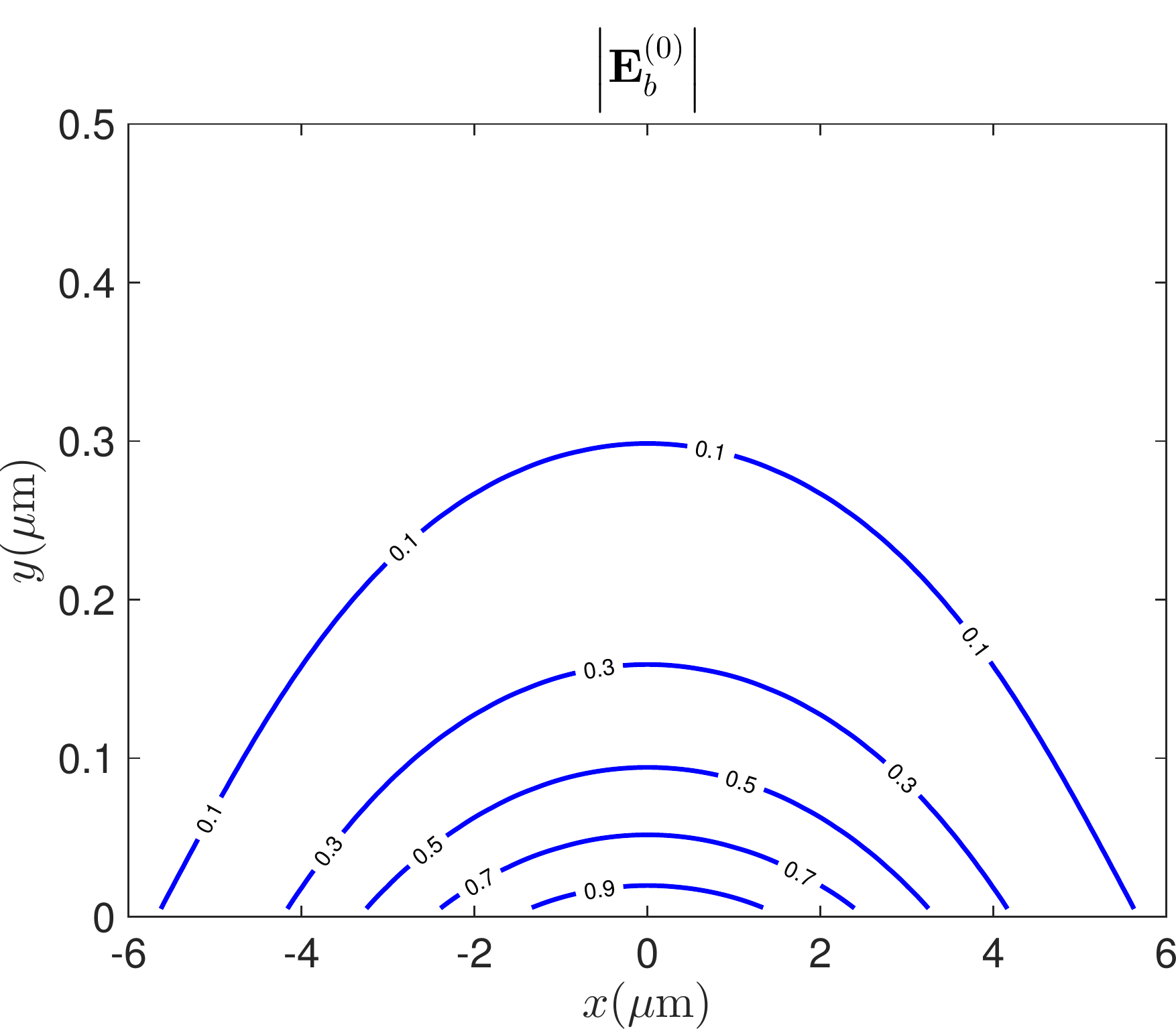}
	\caption{(a) Illustration of laser written and polished waveguide at the surface. 
	Refractive indices are indicated: maximum $n_1+\Delta n$ for the waveguide,
	$n_1$ for unmodified fused silica in the bulk, and $n_0$ for the vacuum.
	(b) Characteristic field profile of the fundamental guiding mode for a laser written surface waveguide,
	centered at $x=0$:
	we note the exponential decay into the air, along $y$ axis (see text for details).}
\label{fig:profile}	
\end{figure}

In this paper we analyze evanescent fields of laser written waveguides, and address the possibility of their application in designing a novel light-matter interface, which is schematically depicted in Figure \ref{fig:profile}(a).
The advantages of our proposed scheme include (a) the benefit from established and developed techniques to manufacture silica glass chips with written waveguides,
(b) the robustness and simplicity of on-chip light-matter interface, and (c) the scalability of the interface, since individual elements can be combined via waveguides or optical fibers.
An integral part of our novel light interface is the ability to position and transport atoms across the chip.
This is why we show here that, for a set of realistic experimental parameters, trapping of atoms can be implemented at distances very close to the surface of a chip with laser written waveguides.
For this we examine the idea of two-color laser trapping and show that a stable trap in two dimensions can be achieved by choosing the geometry of the waveguides and wavelengths of the light such that there are two guided modes for the blue detuned light, while the red detuned light operates on a single guided mode.
In addition to this, we optimize parameters of the structure in order to maximize the trap depth and minimize trap losses for a given input power of light.

The paper is organized as follows: in Section \ref{sec:ev_modes} we introduce the geometry of the waveguide structure written close to the surface of the bulk medium, discuss the eigenmodes of the system \cite{Jukic+:SPIE9379:2015}, and present a simple, analytic model for the evanescent field.
We then construct the total atomic potential for a specific example of Cesium atoms, which consists of the optical potential for the blue and red detuned laser frequencies and an attractive surface potential.
We first demonstrate that trapping is not to be expected when working in a single-mode regime for both colors due to losses to the surface at the sides of the trap.
We then show that a small admixture of a higher blue mode resolves this problem.
Therefore, we restrict ourselves to geometries supporting two blue modes and a single red waveguide mode.
Further, we choose the geometry (within experimental limitations) so that we minimize effective mode area at the surface (for the blue light), essentially maximizing the evanescent field intensity for a given propagating power.
In Section \ref{sec:optim} we study the trap depth and losses as a function of several parameters: distance from the surface, total power of blue and red light, and laser detunings.
In particular, for a given trap depth and total power, we find the detunings for which the trap losses are minimized.
In Section \ref{sec:concl} we summarize our discussion.
\marginpar{!}
%%
%% Come back to that, maybe say that minimsing the mode area is a general concern
%%

%%%
%%% Evanescent field of exposed waveguide%%%

\section{Evanescent field of exposed waveguides}
\label{sec:ev_modes}

We start by describing the structure of a typical laser written waveguide manufactured in the group of Alexander Szameit at the University of Jena \cite{SzameitNolte:JPB43:2010}.
A dielectric medium is exposed to femtosecond Ti:Saphire laser pulses, which creates a permanent refractive index change inside the bulk of the glass.
The shape of the resulting waveguide is generally elliptic in two dimensions, and the dielectric profile of the bulk can be described with a super-Gaussian function \cite{Rechtsman+:NatPhot7:2012},
\begin{equation}
	n(x,y) =  n_1 + \Delta n \exp \left[- \left( \frac{x^2}{r_\mathrm{x}^2} + \frac{(y - d)^2}{r_\mathrm{y}^2} \right)^3 \right].
\label{eq:supergauss}
\end{equation}
Here, $n_1$ is refractive index of unchanged medium, and $\Delta n$ is the relative change in the refractive index.
The ellipse is characterized by two radii $r_\mathrm{x}$ and $r_\mathrm{y}$ and we make no assumption on which is the major and minor radius.
We choose coordinates such that the boundary of the bulk material (containing the waveguide) and the vacuum with refractive index $n_0 =1$ is at $y=0$.
In our setup, waveguides are written in the bulk of fused silica, and then polished to remove the top layer in order to expose the waveguide to the surface.
We assume that the semi-axes are aligned with the axes of the coordinate system. 
An illustration of this laser written and polished waveguide at the surface is plotted in Figure \ref{fig:profile}(a).
In the process of polishing the waveguide is also partially removed; in order to model the resulting geometry we introduce a parameter $d$, which indicates the distance of the semi-axis along $x$ from the polished surface at $y=0$.
Typical length scales for the radii $r_\mathrm{x}$ and $r_\mathrm{y}$ are several microns.

In what follows, we will specify the dielectric medium to be fused silica (SiO$_2$) and that the value of index change is $\Delta n \approx 0.005$, which corresponds to maximum change presently obtained in experiments.
For the range of frequencies we will explore, i.e., close to D1 and D2 resonances of Cesium (with wavelengths $894.59 \ \mbox{nm}$ and $852.12 \ \mbox{nm}$ \cite{Steck:Cs:2010}), we set $n_1 = 1.453$ throughout.

For the numerical calculation of the waveguide eigenmodes we use a vectorial finite difference method as developed in \cite{Fallahkhair+:JLT26:2008}.
In Figure \ref{fig:profile}(b), we show a characteristic profile of the evanescent part of the fundamental HE mode (for the particular example presented, parameters used are $r_\mathrm{x}=5$ \textmu m, $r_\mathrm{y} = 4$ \textmu m, and the light is blue detuned $10 \ \mbox{nm}$ from the D2 resonance).
\marginpar{!}
%%
%% Atomic spectrum has not been introduced yet ...
%%

For a large set of parameters, the dominant field component of the fundamental waveguide solution in the evanescent region can simply be modeled as,
\begin{equation}
	E_a^{(0)}(x,y) \approx A_a^{(0)} \ \rme^{-y/y_a} \rme^{-x^2/x_a^2},
\label{eq:field_approx}  
\end{equation}
where $a = b,r$ refers to the "color" of the detuning, and $x_\mathrm{b,r}$ are characteristic width for blue and red detuned waveguide modes. 
The decay lengths $y_\mathrm{b,r}$ are given by
\begin{equation}
	y_a = \frac{\lambda_a}{2\pi} \left( n^2_a - n_0^2 \right)^{-\frac{1}{2}},
	\label{eq:decaylength}
\end{equation}
where $a = b,r$ and $n_a$ is the effective refractive index obtained from the numerical solution of the given geometry for respective blue or red detuned wavelength.
The characteristic widths $x_{b,r}$ are correlated to the radius $r_\mathrm{x}$ of the waveguide.
\marginpar{!}
%%
%% effective refractive index not introduced, show link between radius and width?
%% explain fitting of numerial date to Gaussians
%%

%The accuracy of this simple model if demonstrated in Figure \ref{fig:gaussian}, where we compare the Gaussian profile at the interface for $y = 0$ and the evanescent behavior for $x=0$ with the exact numerical solution.
We will use our simple model of the evanescent field to derive general statements aboout the requirements on the widths $x_\mathrm{b}$ and $x_\mathrm{r}$ (and therefore on the radius $r_x$), as well as the decay lengths $y_\mathrm{b,r}$ to form a stable trap.
However, all final results are based on the full numerical calculation of the guided modes.
%
% \begin{figure}
%  	\includegraphics[width=0.45\textwidth]{}
%  	\includegraphics[width=0.45\textwidth]{}
% % 	\includegraphics[width=0.8\textwidth]{Presentation1.eps}
% 	\caption{(a) Illustration of laser written and polished waveguide at the surface.
% 	Refractive indices are indicated: maximum $n_1+\Delta n$ for the waveguide,
% 	$n_1$ for unmodified fused silica in the bulk, and $n_0$ for the vacuum.
% 	(b) Characteristic field profile of the fundamental guiding mode for a laser written surface waveguide,
% 	centered at $x=0$:
% 	we note the exponential decay into the air, along $y$ axis (see text for details).}
% \label{fig:gaussian}
% \end{figure}
%
Varying the parameter $d$ has a similar effect on the eveanescent field as changing the radius $r_\mathrm{y}$.
This is why we set $d=0$ from now on for concreteness, meaning that exactly half of the waveguide is cut.

Recently, we have presented more detailed discussion of fundamental modes in surface laser written waveguides, together with the resulting evanescent part reaching into the vacuum \cite{Jukic+:SPIE9379:2015}.
The waveguide propagating along $z$ axis can have two quasi-degenerate hybrid solutions for the fundamental guiding mode.
The first solution (we label it as HE) has electric field with dominant polarization along $x$ axis, and the second (EH) solution along $y$ axis.
However, in the evanescent part, the second solution has a significant longitudinal $z$ component of the  
electric field which is comparable to the $y$ component.
This suggests that the EH mode has in general elliptic polarization in evanescent region, and carries spin angular momentum orthogonal to the propagation direction \cite{Aiello+:PRL103:2009,DennisGoette:JOpt15:2013,Bliokh+:NatComm5:2014}.
In order to avoid possible non-scalar light shift contributions, in what follows we will assume only HE solutions (quasi-$x$ polarized) for both blue and red detuned light.

%%%
%%% Trapping potential
%%%

\section{Trapping potential}

The basic idea behind trapping of atoms in evanescent field of a waveguide is to use two different lasers which are blue and red detuned from the electronic transition resonances \cite{Balykin+:PRA70:2004}.
The blue (red) detuned color leads to a repulsive (attractive) optical potential which is proportional to the intensity of light.
For these forces the total scalar contribution to the light shift is,
\begin{equation}
  V_\mathrm{light}(x,y)=-\frac{1}{4} \alpha_\mathrm{b} \left| \mathbf{E}_b(x,y) \right|^2 -\frac{1}{4} \alpha_\mathrm{r} \left| \mathbf{E}_r (x,y) \right|^2 ,
\end{equation}
where $\alpha_\mathrm{b,r}$ are the frequency dependent real polarizabilities of the Cs ground state for the blue and red detuned laser fields $\mathbf{E}_{b,r}$ \cite{Grimm+:AdvAMOPhys42:2000, LeKien+:EPJD67:2013}.
We assume throughout the paper that the two lasers are blue detuned from the D2 resonance of Cs, 
and red detuned from the D1 resonance.

However, in addition to the optical potential we have to include the effect of attractive surface forces due to van der Waals and Casimir-Polder interactions.
For the Cs atomic trap at the distance of a few $100 \ \mbox{nm}$ above the interface, the surface potential can be approximated by the heuristic form,
\begin{equation}
  V_\mathrm{surf} (y) = - \frac{C_4}{C_3} \frac{1}{(C_3 y + C_4)y^3},
\end{equation}
with $C_3(6S_{1/2})/h=1.16$ \mbox{kHz} \textmu $\mbox{m}^3$, and $C_4(6S_{1/2})/h=0.15$ \mbox{kHz} \textmu $\mbox{m}^4$ \cite{Lacroute+:NJP14:2012, Stern+:NJP13:2011}.
The total potential in the $x,y$ plane is therefore%
\begin{equation}
 V(x,y) = V_\mathrm{light}(x,y) + V_\mathrm{surf}(y).
\end{equation}
%
%In the following we examine conditions on the parameters characterizing the light field in order to generate a stable trapping minimum above surface of the waveguide.

% \subsection{Limitations of atom trapping in single-mode regime}
%
% Previous proposals and experimental verification of two-color trapping schemes in dielectric nanofibers have relied on single waveguide mode for both the blue and red detuned lasers \cite{Balykin+:PRA70:2004, LeKien+:PRA70:2004, Vetsch+:PRL104:2010}.
% The dielectric nanofiber supports only the fundamental modes for both blue and red light, and their superposition is sufficient to form a stable atom trap.

If we neglect the surface potential for the moment and make use of the Gaussian approximation if Eq. (\ref{eq:field_approx}) we can write the optical potential as 
\begin{equation}
	\fl V_\mathrm{light} (x,y) \approx - \frac{\alpha_\mathrm{r}}{4} \left( A_\mathrm{r}^{(0)} \right)^2 \ \rme^{-2 y/y_\mathrm{r}} \rme^{-2 x^2/x_\mathrm{r}^2} - \frac{\alpha_\mathrm{b}}{4} \left( A_\mathrm{b}^{(0)} \right)^2 \ \rme^{-2 y/y_\mathrm{b}} \rme^{-2 x^2/x_\mathrm{b}^2},
\end{equation}
which, on introducing scaled, dimensionless variables $\xi = x/x_\mathrm{r}$ and $\upsilon = y / y_\mathrm{r}$ can be fully characterised by the ratios of the amplitudes $\tilde{A}_\mathrm{b/r} = A_\mathrm{b}^{(0)}/A_\mathrm{r}^{(0)}$, the ratio of the polarizabilities $\tilde{\alpha}_\mathrm{b/r} = \alpha_b / \alpha_r$, the ratio of the ratio of the decay lengths $\tilde{y}_\mathrm{r/b} = y_\mathrm{r} / y_\mathrm{b}$, and the Gaussian widths $\tilde{x}_\mathrm{r/b} = x_\mathrm{r} / x_\mathrm{b}$.
The resulting form of the optical potential is then given by
\begin{equation}
	\label{eq:potential_scaled}
	\fl V_\mathrm{light} (\xi,\upsilon) \approx - \frac{\alpha_r}{4} \left( A_\mathrm{r}^{(0)} \right)^2 \ \rme^{-2 \upsilon} \rme^{-2 \xi^2} \left[ 1 + \tilde{\alpha}_\mathrm{b/r}\left( \tilde{A}_\mathrm{b/r} \right)^2 \rme^{-2 \upsilon ( \tilde{y}_\mathrm{r/b} - 1 )} \rme^{-2 \xi^2 ( \tilde{x}^2_\mathrm{r/b} - 1 )} \right]
\end{equation}
The shape of the trapping potential is determined by the expression in the brackets.
This is becomes most obvious when first considering the conditions for a global trapping minimum at $\xi = 0$.
Determining the position $\upsilon_0$ of an extremum of Eq. (\ref{eq:potential_scaled}) along $\xi=0$ gives rise to the following condition:
\begin{equation}
	\left[ 1 + \tilde{y}_\mathrm{r/b} \tilde{\alpha}_\mathrm{b/r}\left( \tilde{A}_\mathrm{b/r} \right)^2 \rme^{-2 \upsilon_0 ( \tilde{y}_\mathrm{r/b} - 1 )} \right] = 0.
\end{equation}
Demanding that this extremum is a minimum along both $\xi$ and $\upsilon$ and not a saddle point leads to the two additional conditions
\begin{eqnarray}
	\left[ 1 + \tilde{y}^2_\mathrm{r/b} \tilde{\alpha}_\mathrm{b/r}\left( \tilde{A}_\mathrm{b/r} \right)^2 \rme^{-2 \upsilon_0 ( \tilde{y}_\mathrm{r/b} - 1 )} \right] & < & 0, \\
	\left[ 1 + \tilde{x}^2_\mathrm{r/b} \tilde{\alpha}_\mathrm{b/r}\left( \tilde{A}_\mathrm{b/r} \right)^2 \rme^{-2 \upsilon_0 ( \tilde{y}_\mathrm{r/b} - 1 )} \right] & > & 0.
\end{eqnarray}
This in turn sets a condition on the ratios of the decay lenghts and Gaussian widths, as $\tilde{y}_\mathrm{r/b} > 1$ and $\tilde{y}_\mathrm{r/b} > \tilde{x}^2_\mathrm{r/b}$ to fulfill both equations simultaneously. 

%In this form it is obvious that the shape of the trapping potential is determined by the expression in the brackets.
% Choosing the wavelengths of the blue and red detuned lasers, fixes the atomic polarisabilities and it is then a matter of the the appropriate ratio of the amplitudes $\tilde{A}_\mathrm{r/b}$ to ensure that the whole expression in the brackets changes from negative to positive values for varying $\upsilon$ creating a 1D minimum of $V_\mathrm{light}(\xi=0,\upsilon)$ along the $\upsilon$ ot $y$ axis.
The properties of this trap are then determined by the ratios of the decay lengths $\tilde{y}_\mathrm{r/b}$ and the Gaussian widths $\tilde{x}_\mathrm{r/b}$.
As the effective refractive index $n_\mathrm{eff}$ does not change drastically for different wavelengths, according to Eq. (\ref{eq:decaylength}) the former is approximatively given by the ratio of the wavelengths $\tilde{y}_\mathrm{r/b} \approx \lambda_\mathrm{r} / \lambda_\mathrm{b}$, which therefore satisfies $\tilde{y}_\mathrm{r/b} > 1$.
The condition for a stable mininum is therefore direclty given by
\begin{equation}
	\label{eq:condition2d}
	\tilde{y}_\mathrm{r/b} > \tilde{x}^2_\mathrm{r/b}.
\end{equation}	
The dependence of the Gaussian widths on the geometry of the waveguide as defined by the radii $r_x$ and $r_y$ is more complicated.
However, we are aiming to create a trap close to the surface ($\sim 100 \ \mbox{nm}$ at most) with waveguide modes having a small effective area.
% Whereas the former is approximatively given by the ratio of the wavelengths $\tilde{y}_\mathrm{r/b} \approx \lambda_\mathrm{r} / \lambda_\mathrm{b}$ according to Eq. (\ref{eq:decaylength}), as the effective refractive index $n_\mathrm{eff}$ does not change drastically for different wavelengths, the dependence of the Gaussian widths on the geometry of the waveguide as defined by the radii $r_x$ and $r_y$ is more complicated (see Fig. \ref{fig:wg_geometry} a).
% However, we are aiming to create a trap close to the surface ($\sim 100 \ \mbox{nm}$ at most) with waveguide modes having a small effective area.
This is why we are concentrating on geometries, for which the waveguide size defined by $r_x$ is a few microns and commensurate with the wavelength.
In this parameter range the ratio of the resulting Gaussian widths also scales approximatively with the ratio of the wavelengths $\tilde{x}_\mathrm{r/b} \approx \lambda_\mathrm{r} / \lambda_\mathrm{b}$. 
As as consequence it is not possible to form a stable trap in this parameter regime.

% As as consequence it is not possible to form a stable trap in this parameter regime, as the requirement for a full confinement in two dimensions is given by
% %
% \begin{figure}[ht]
% \includegraphics[width=0.45\columnwidth]{}
% \caption{Geometry}
%
% \label{fig:wg_geometry}
% \end{figure}
% %
% \begin{equation}
%   \frac{y_\mathrm{r}}{y_\mathrm{b}} > \left( \frac{x_\mathrm{r}}{x_\mathrm{b}} \right)^2,
%   \label{condition2d}
% \end{equation}
% %
% which is derived from the condition that the curvature of the optical potential, and hence the second order derivative, needs to be positive in both directions to form a minimum in two dimensions.
% Otherwise, the stationary point is only a saddle point, with no confinement along the $x$ axis.
% To make the constraint Eq. (\ref{condition2d}) more transparent, in the limit when $\frac{y_\mathrm{r}}{y_\mathrm{b}} \approx \frac{x_\mathrm{r}}{x_\mathrm{b}}$ the saddle point can equivalently be explained by a different sign of a second order partial derivative at the stationary point calculated with respect to $x$ coordinate (Gaussian) and the one calculated with respect to $y$ coordinate (exponential function).

\begin{figure}[ht]
\includegraphics[width=0.45\columnwidth]{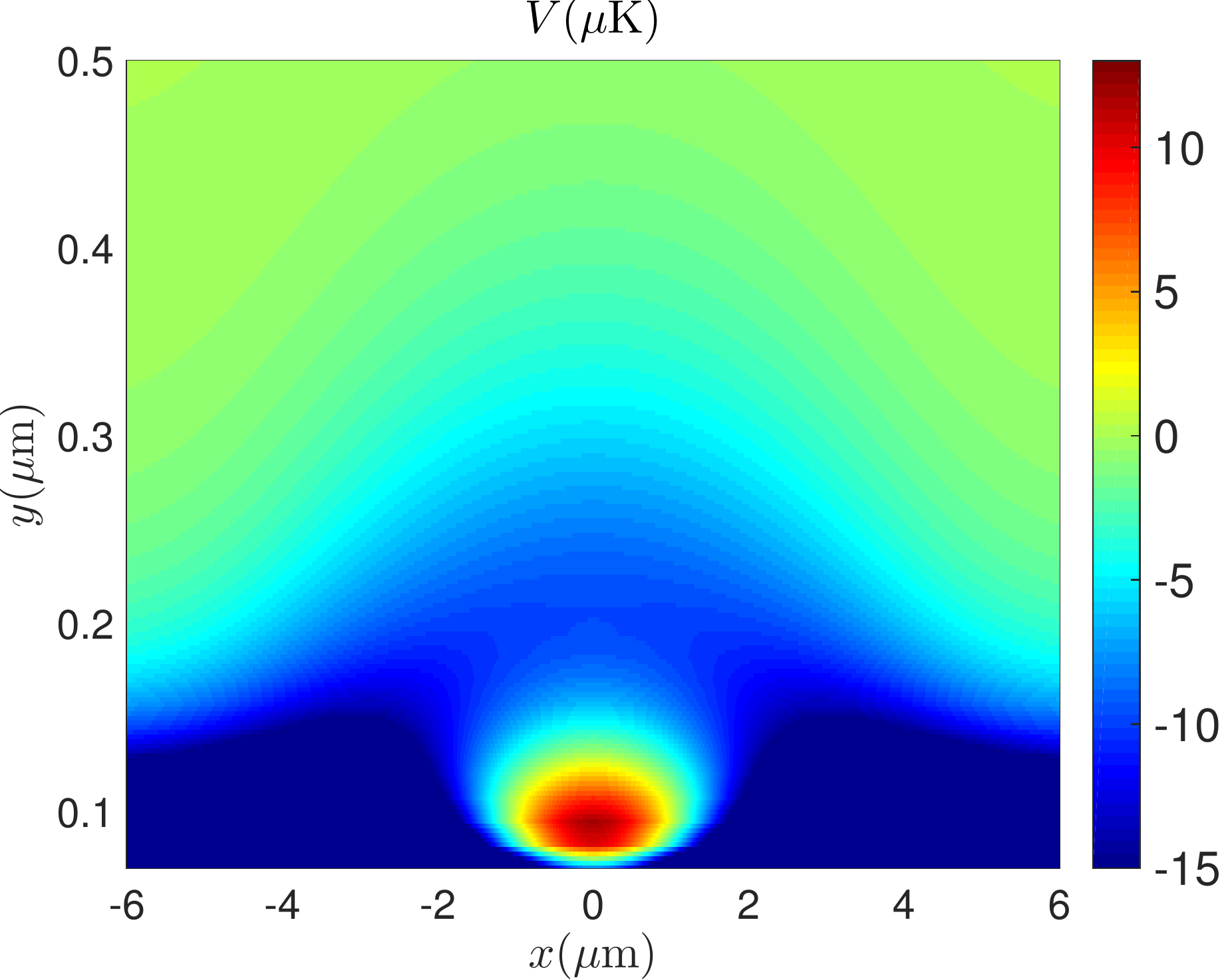}
\caption{Total 2D potential for Cs atoms created by superimposing single-mode blue and red detuned lasers, together with attractive surface force.
Here, the saddle point is at $\sim 200\ \mbox{nm}$ above the surface.
}
\label{fig:saddle_point}
\end{figure}

In general, condition Eq. (\ref{eq:condition2d}) can be satisfied in certain limits, for example when the waveguide size becomes large compared to the wavelength.
In this case, the resulting widths $x_\mathrm{b}$ and $x_\mathrm{r}$ of the Gaussians along $x$ are not only given by the wavelengths, but depend also on the radii $r_\mathrm{x}$ and $r_\mathrm{y}$.
However, the resulting optical potential along $x$ axis is then extremely shallow.
In addition to this, the presence of attractive surface forces makes the trapping along $x$ axis even more constrained.
We have numerically verified that for realistic experimental parameters and working only with the fundamental modes, which even obey equation Eq. (\ref{eq:condition2d}), the resulting trapping potential along $x$ is at least one order of magnitude smaller then the potential along the $y$ direction, which is problematic as it can therefore also be smaller than the  characteristic harmonic oscillator energy of the trap along the $y$ direction.
A typical example of the total atomic potential with a saddle point is presented in Figure \ref{fig:saddle_point}.
Here, the saddle point is located at about $\sim 200\ \mbox{nm}$ from the surface.
For this example the parameters used are $r_\mathrm{x}=5$ \textmu m, $r_\mathrm{y} = 4$ \textmu m, and the blue and red light are detuned $10 \ \mbox{nm}$ from respective resonances with a combined input power of $8$ W.

\subsection{Atomic trap with two blue modes}

\begin{figure}[ht]
\includegraphics[width=0.45\columnwidth]{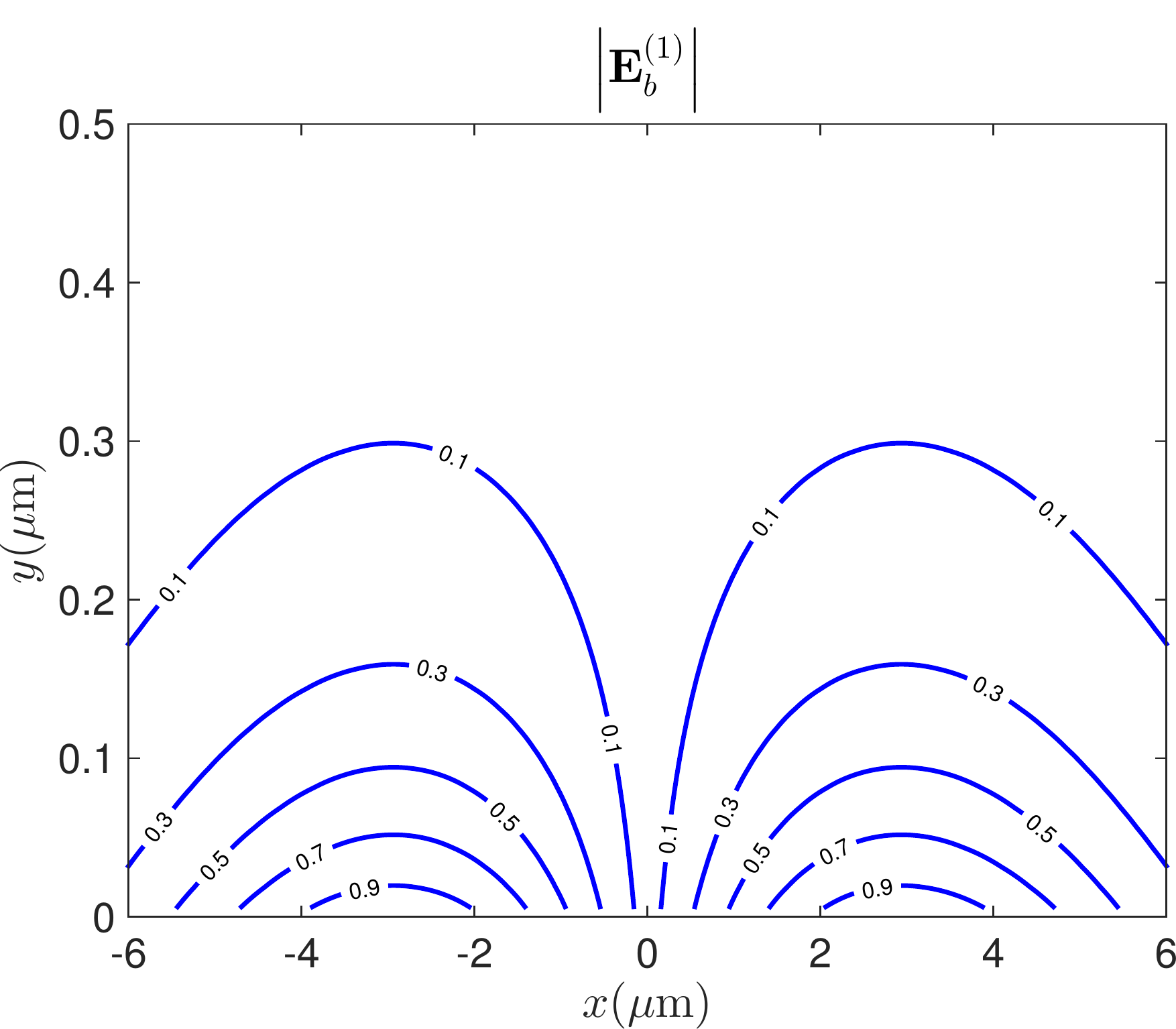}
\includegraphics[width=0.45\columnwidth]{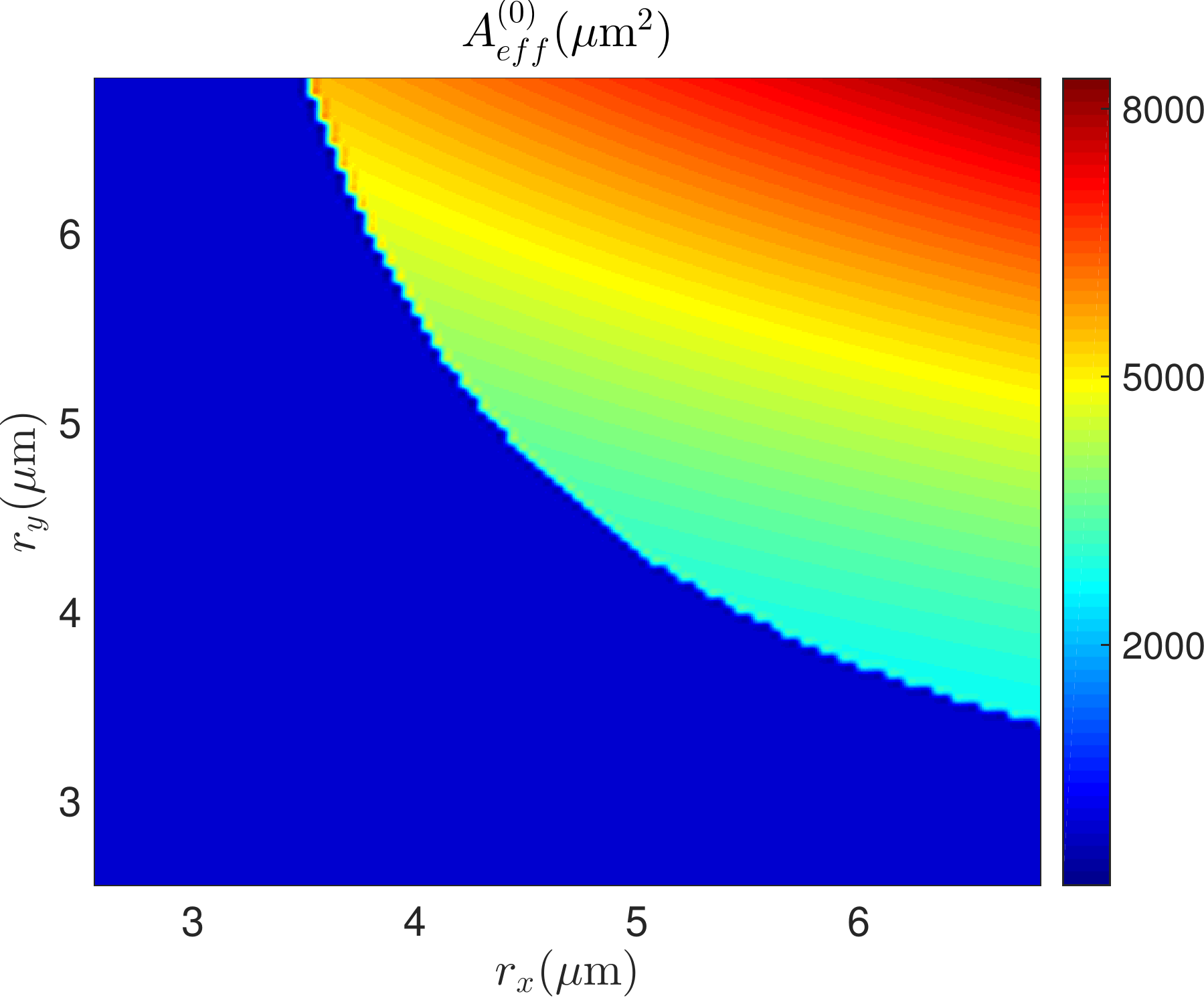}
\caption{(a) Upper waveguide mode of blue detuned light with two intensity maxima that can provide for 2D atomic confinement.
(Waveguide parameters are $r_\mathrm{x}=6$ \textmu m, $r_\mathrm{y} = 3.9$ \textmu m, and the light is blue detuned $10$ nm from D2 resonance).
(b) Effective area at the surface  $A_\mathrm{eff}^{(0)}$ for the resonant wavelength $\lambda_{D2}$ of the fundamental guiding mode. 
Blue region is omitted from the plot since it does not support upper waveguide mode, and therefore cannot be used for the design of efficient 2D atomic trap.
}
\label{fig:2nd_mode_and_mode_area_of_low_lamdba_b}
\end{figure}

In this subsection we show that full 2D atomic confinement can be efficiently created by introducing an upper mode contribution to the blue detuned light.
The general idea is that two intensity maxima to the left and right from the symmetry axis $x=0$, characteristic for a higher waveguide mode, can create a repulsive potential at the sides of the stationary point that is large enough to turn the saddle point of atomic potential into a local minimum.
A characteristic field profile of the upper mode is shown in Figure \ref{fig:2nd_mode_and_mode_area_of_low_lamdba_b}(a).
%
%
% \begin{figure}[ht]
% \includegraphics[width=0.45\columnwidth]{upper_mode.eps}
% \caption{Upper mode.
% }
% \label{fig:upper_mode}
% \end{figure}
% %
%
Obviously, for this trapping mechanism to be implemented, it is necessary to choose waveguide geometries that support not only the fundamental, but also a higher (second) mode for the blue detuned light.
To avoid the possibility of coupling into a higher order mode for the red detuned light, we make use of the fact that the red detuned has a longer wavelengths, rendering it possible to design a waveguide supporting a single red and two blue detuned modes.

Under the general premise to create a working trapping devive which strive to optimize the geometry.
This is even more important now as adding another mode increases the parameter space.
One way to optimize the trapping scheme is maximize the electric field density of the fundamental (blue) mode at the surface, $| \mathbf{E}_{b,0}^{(0)} |^2$, for a given total propagating power $P_{b}^{(0)}$ of this mode.
For this we define effective area of the fundamental blue mode at the surface point $(x,y)=(0,0)$ as,
\begin{equation}
 A_\mathrm{eff}^{(0)}= \frac{| \mathbf{E}_{b,0}^{(0)} |^2}{P_{b}^{(0)}}.
\end{equation}
In other words, we require that the effective area of the fundamental mode is minimized.
%The idea is illustrated in Fig. \ref{fig:2nd_mode_and_mode_area_of_low_lamdba_b}.
In Figure \ref{fig:2nd_mode_and_mode_area_of_low_lamdba_b} we have plotted the effective area of the fundamental mode at the D2 resonance as a function of waveguide geometry. 
We have also excluded geometries which do not support two-mode guiding (blue colored region).
%%
%% figure comparing eff area and width of trap
%%
%

From this we notice that the optimization requires the use of geometries with larger $r_\mathrm{x}$ and smaller $r_\mathrm{y}$ in the region of the parameter space where two blue modes are supported.
This would suggest that the trap optimized this way might be fairly large in $x$-direction.
In order to avoid possible experimental difficulties with very elongated waveguides, we limit the geometry to $r_\mathrm{x}=6.0$ \textmu m.
Further, we choose $r_\mathrm{y} = 3.9$ \textmu m, thereby enabling two-mode regime for blue detuned lasers, and single-mode regime for red light.

After settling for a geometry, we construct the blue input light field as linear superposition of the fundamental and the upper mode:
\begin{equation}
  \mathbf{E_b}=\sqrt{\tau} \mathbf{E}_{b}^{(0)} + \rmi \sqrt{1-\tau} \mathbf{E}_{b}^{(1)},
  \label{eq:twomode}
\end{equation}
where $\tau$ parametrizes the contributions of the two blue modes.
The phase difference of $\pi/2$ is important, as the higher mode $\mathbf{E}_{b}^{(1)}$ has as $\pi$ phase jump across the symmetry axis in its dominant component.
In this the phase difference between the fundamental and higher mode is the same, albeit with a different sign, which nevertheless gives rise to a symmetric intensity distribution.
\marginpar{!}
%%
%% how to generate phase difference?
%%

The two blue modes have slightly different propagation constants or effective refractive indices $n_b^{0}$ and $n_b^{(1)}$.
This will create a beating of the two modes which changes the phase relation in Eq. (\ref{eq:twomode}) and therefore the shape of the trapping potential. 
The beating period, given by $\lambda_b/(n_b^{0} - n_b^{1})$, is, however, very long, owing to the small difference in the effective refractive indices.
We can give a strict upper bound for the beating length by noticing that the maximum difference in the effective refractice index is given by the contrast $\Delta n$, which would correspond to a period of 200 $\lambda_b$, which is about $170$ \textmu m. 
More realistically, the difference between the two effective refractive indices is about $0.001$ we gives a beating period of $1000 \lambda_b$ or $850$ \textmu m.
To generate a full 3D confinement of the atoms it may therefore be necessary to reflect the field in Eq. (\ref{eq:twomode}) to create a standing wave modulated by the beating frequency.
\marginpar{!}
%%
%% explain futher, final 3d trapping volume
%%

\begin{figure}[ht]
\includegraphics[width=0.45\columnwidth]{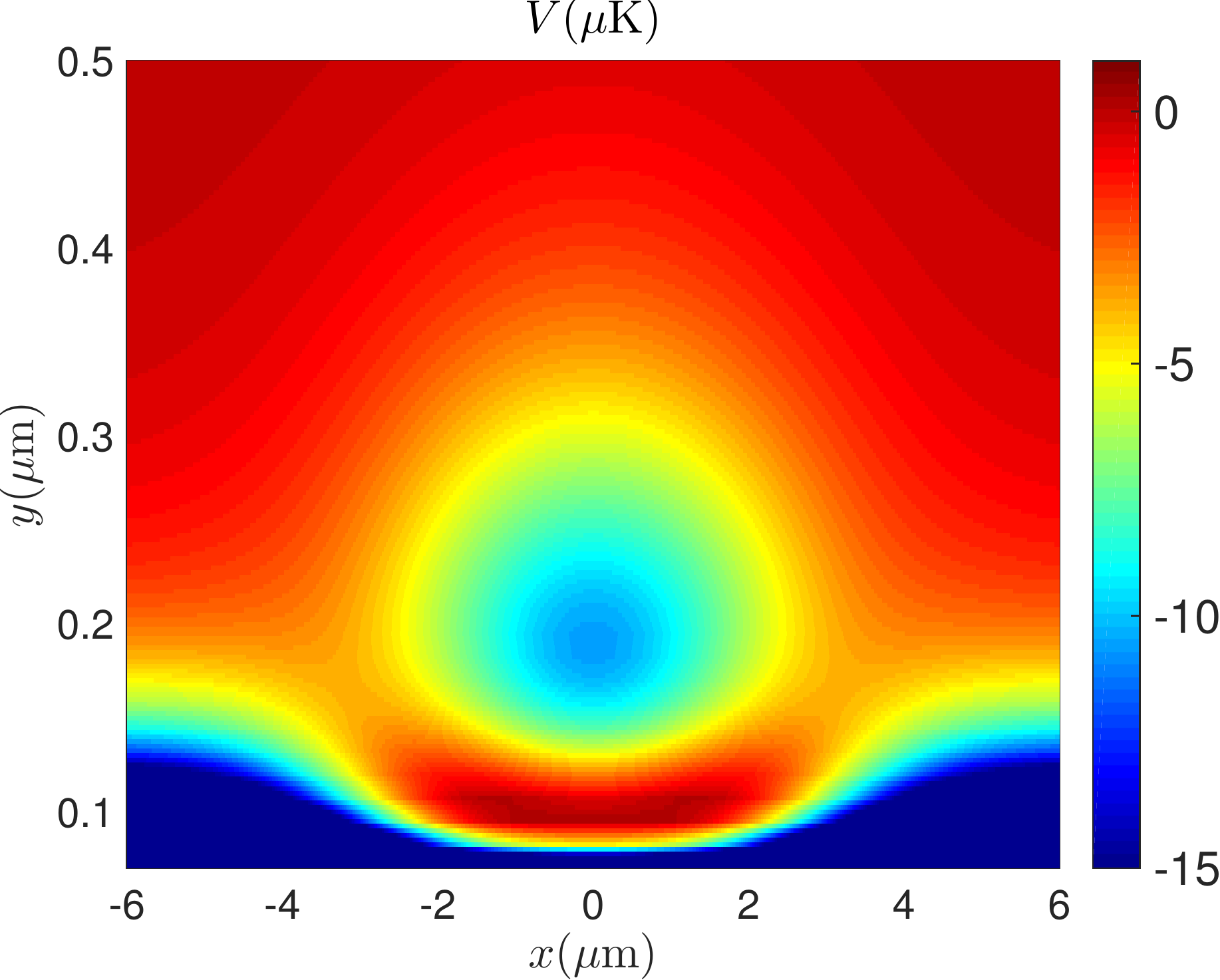}
\includegraphics[width=0.45\columnwidth]{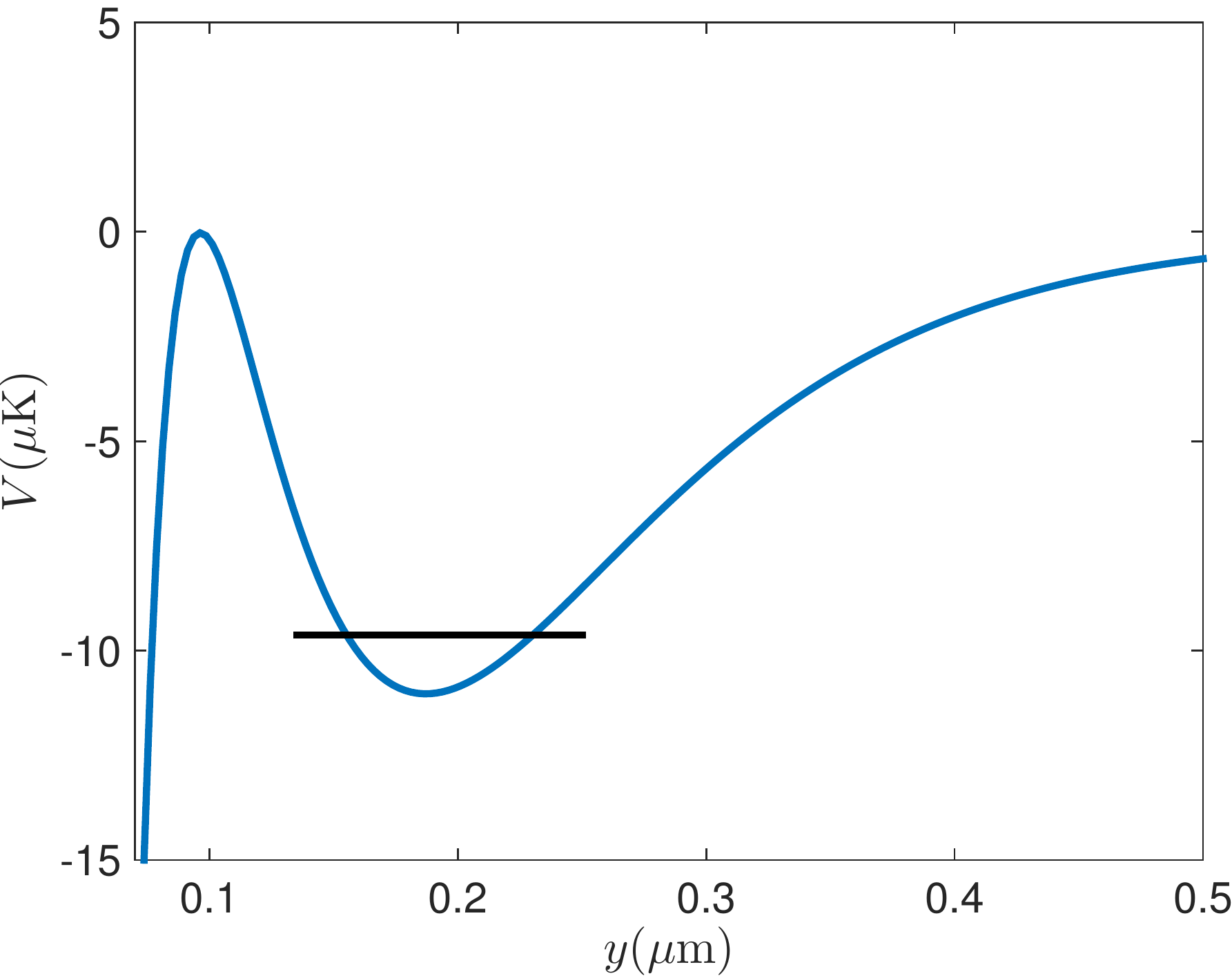}
\caption{(a) Two-mode trap for total power $P_\mathrm{in} = 8$ W, with $P_\mathrm{b} = 3.66$ W for blue, and $P_\mathrm{r} = 4.34$ W, for red detuned light.
Detunings from D1 and D2 resonances are $\delta \lambda_\mathrm{r} = 10.5$ nm, and $\delta \lambda_\mathrm{b} = 10$ nm. 
(b) Two-mode trap along $x=0$ line; the black like denotes energy of the trap ground state energy in harmonic approximation.}
\label{fig:two_mode_trap}
\end{figure}

Even a small presence of the upper mode, relative to total input power, can produce 2D confinement.
Therefore, in what follows, we set $k=0.95$.
In Figure \ref{fig:two_mode_trap} we present an example of 2D atomic trap.
It clearly demonstrates that the trap stability can be achieved by including upper blue mode for trapping.
For this example we assume the total power propagating along the waveguide to be $8$ W,
since the laser written waveguides can easily cope with input powers of several Watts without modifying their physical properties.
In general, we can tune depth, width, and position of the trap minimum (and also losses), by changing waveguide geometry, total input power of lasers, contributions of blue and red light fields, and their detunings.
We note that the characteristic length scale of the trap in $x$-direction is more then one order of magnitude larger then along $y$ axis, i.e. for the corresponding oscillator frequencies of the trap we have $\omega_x \ll \omega_y$,
so that the ground state energy is $E \approx E_y = \frac{\hbar \omega_y}{2}$.
Therefore, we can focus on a study of one-dimensional potential along $x=0$ line illustrated at Figure \ref{fig:two_mode_trap}(b).
In particular, we use the 1D profile to define trap depth relative to both the potential barrier close to the surface and to the zero potential away from the surface.

\begin{figure}[ht]
\includegraphics[width=0.45\columnwidth]{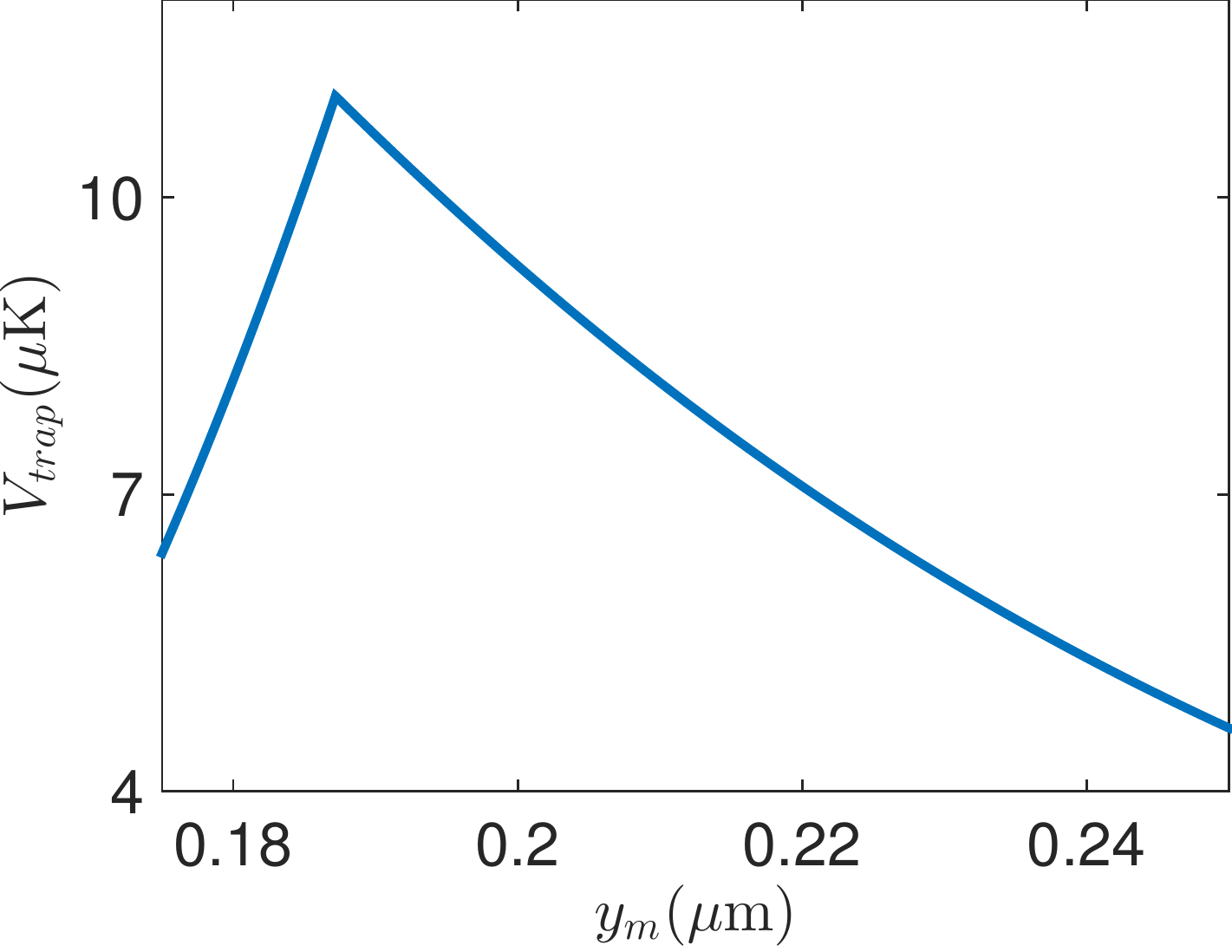}
\includegraphics[width=0.45\columnwidth]{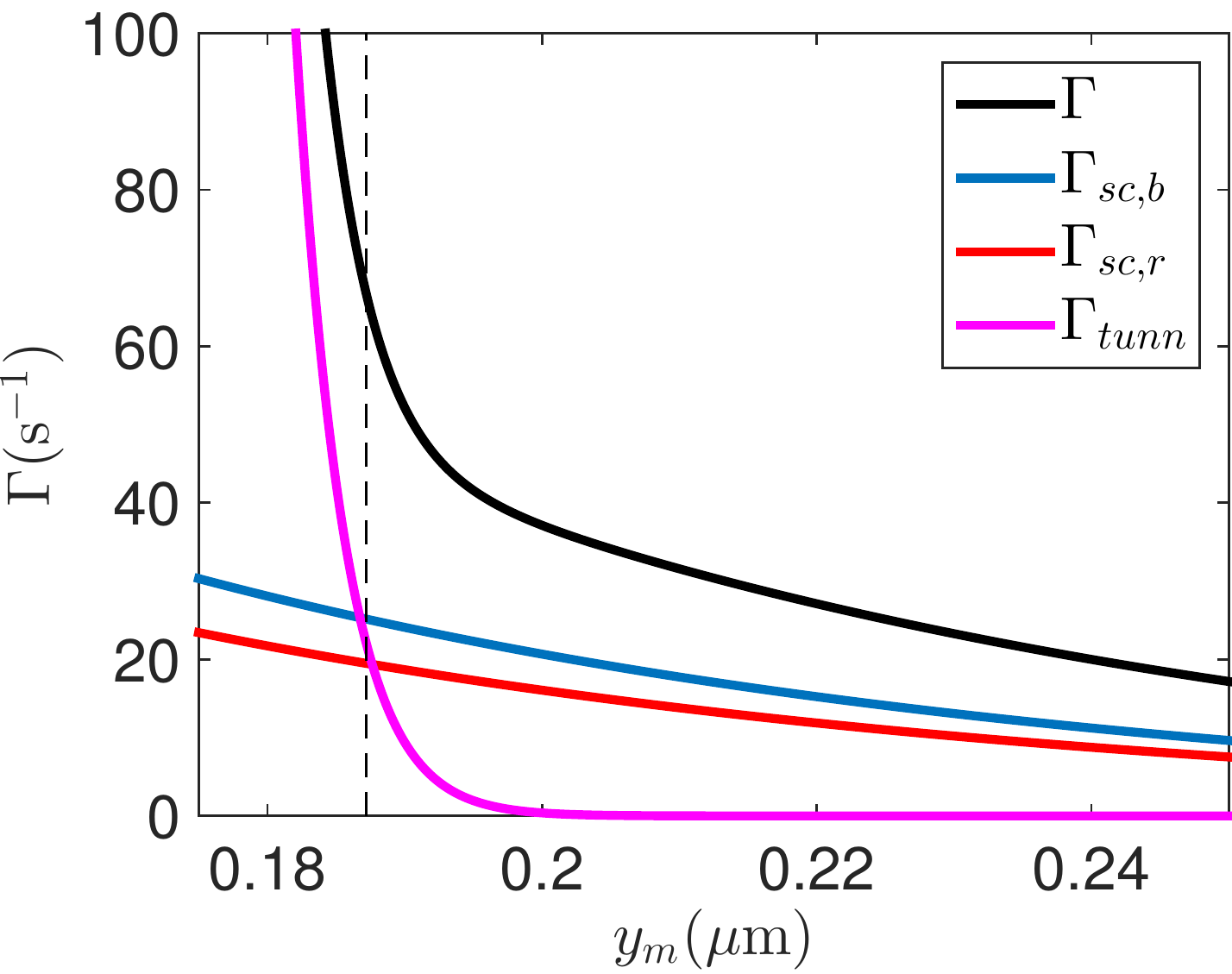}
\caption{
(a) Trap depth as a function of the position of the trap minimum.
Total input power and detunings are the same as in Figure \ref{fig:two_mode_trap}.
(b) Losses as a function of position of the trap minimum.
Dashed vertical line indicates the position with the maximum trapping potential, i.e., example from Fig. \ref{fig:two_mode_trap}.
%Horizontal line in (a) presents WKB calculated tunnelling loss $\Gamma_{tunn} \approx 27.4 \ \mbox{s}^{-1}$ for this example.
Total loss rate is then $\Gamma \approx 67 \ \mbox{s}^{-1}$.
}\label{fig:losses_vs_ym}
\end{figure}

The specific example presented in Figure \ref{fig:two_mode_trap} corresponds to the maximum trap depth ($\sim 11$ \textmu K) for the given choice of total power and the detunings.
Namely, for the set of fixed system parameters (geometry, total input power and detunings),
the trap depth and position depend only on relative contributions of blue and red detuned laser intensities.
Therefore, we can explore the depth of the trap as a function of its location $y_\mathrm{min}$ as shown in Figure \ref{fig:losses_vs_ym}(a).
We notice that the trap becomes more shallow for large distances $y_\mathrm{m}$:
we explain this by the fact that optical fields creating the potential decrease further away from surface.
Further, in the regime where distances from the surface are small,
repulsive barrier towards surface becomes also smaller due to the attraction of surface forces,
effectively reducing the trap depth.

\section{Optimization of losses of the atomic trap}
\label{sec:optim}

In this section we want to estimate losses from the atomic trap. We calculate two major contributions: the tunneling to the surface, and the losses due to photon scattering.
We estimate losses to the surface within the WKB approximation along the 1D line $x=0$.
That is, given the oscillator ground state frequency $\omega_y$, we have%
\begin{equation}
 \Gamma_\mathrm{tunn} = \frac{\omega_y}{2 \pi} \exp \left(-2 \int_{y_1}^{y_2} dy \sqrt{\frac{2 m}{\hbar^2} \left[ V(x=0,y) - E \right]} \right).
 \label{gamma_tunn2}
\end{equation}
where $y_{1,2}$ are the turning points of the potential barrier, $V(x=0,y_{1,2}) = E$.
Further, we calculate losses due to photon scattering from blue and red detuned light field, $\Gamma_{sc,b}$ and $\Gamma_{sc,r}$, by extracting both field densities at the center of the trap \cite{Grimm+:AdvAMOPhys42:2000}.
 
\begin{figure}[ht]
\includegraphics[width=0.45\columnwidth]{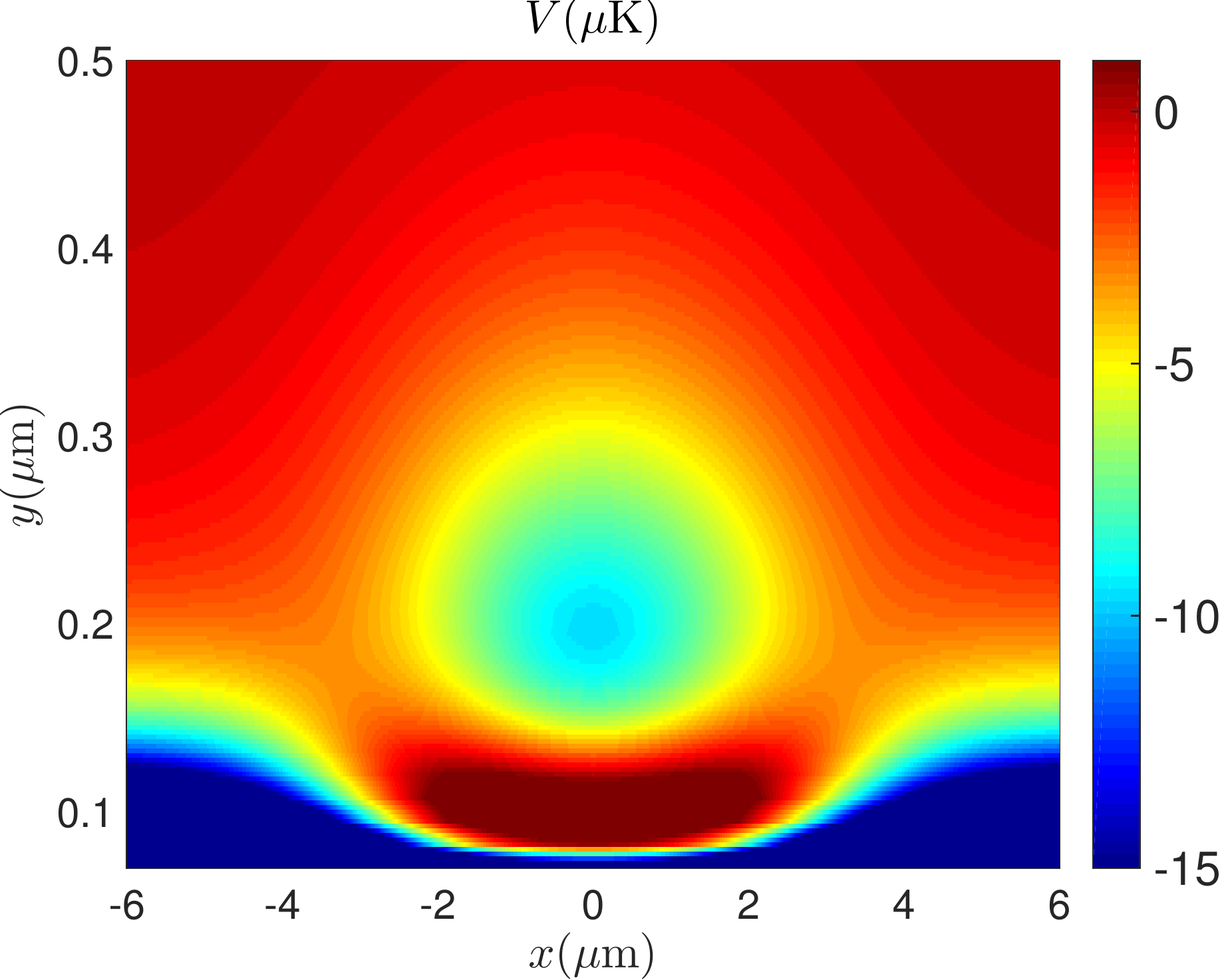}
\includegraphics[width=0.45\columnwidth]{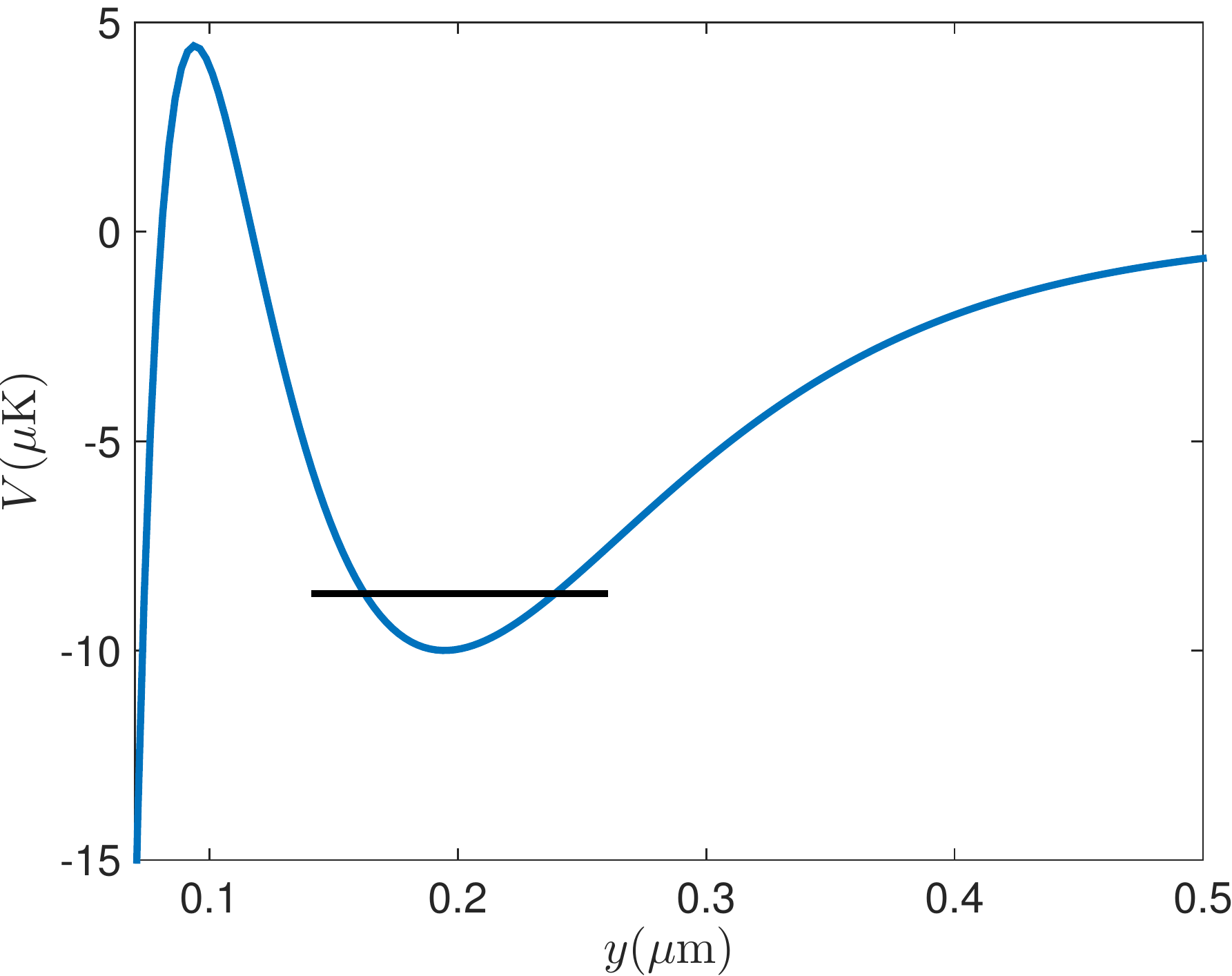}
\caption{(a) Two-mode trap optimized for losses for total power $P_\mathrm{in} = 8$W and $V_\mathrm{trap} = 10$ \textmu K.
Here, we find $\Gamma \approx 42 \ \mbox{s}^{-1}$.
Detunings from D1 and D2 resonances are $\delta \lambda_\mathrm{r} = 10.5$nm, and $\delta \lambda_\mathrm{b} = 10$nm. 
(b) Two-mode trap along $x=0$ line; black line corresponds to the ground state energy of harmonic oscillator.}
\label{fig:two_mode_trap_opt}
\end{figure}

We now proceed to explore possibility of minimizing trap losses by changing trap parameters.
In Fig. \ref{fig:losses_vs_ym}(b) we plot all losses as a function of the position of the trap minimum.
The parameters are identical to those for Figure \ref{fig:losses_vs_ym}(b). 
As expected, we notice that losses increase as the trap location approaches the surface. 
For large trap distances from the surface, the tunneling loss $\Gamma_\mathrm{tunn}$ becomes negligible; however, is increases rapidly relative to the blue and red scattering losses for smaller $y_\mathrm{m}$.
Again, this can be explained by noting that for traps which are too close to the surface, the repulsive potential barrier become too small for efficient atom trapping.
The vertical dashed line corresponds to the maximum trap depth presented in Fig. \ref{fig:two_mode_trap}.
Total loss here is $\Gamma \approx 67 \ \mbox{s}^{-1}$, and we notice that the tunneling loss becomes comparable to the scattering contributions. 
Loss considerations can therefore lead to modified trap optimization strategy.
In other words, we can reduce losses by pushing the trap minimum further away from the surface at the expense of having the trap also more shallow.
%
%
%This means that the trap properties can be made better by slightly more shallow trap.
Example of such trap with $V_\mathrm{trap} =  10$ \textmu K, and $\Gamma \approx 42 \ \mbox{s}^{-1}$, is shown in Fig. \ref{fig:two_mode_trap_opt}.
As can be observed, the barrier to the surface is enlarged, effectively leading to very small tunneling loss.
This example can also be considered a result of loss optimization process described below.

\begin{figure}[ht]
\includegraphics[width=0.45\columnwidth]{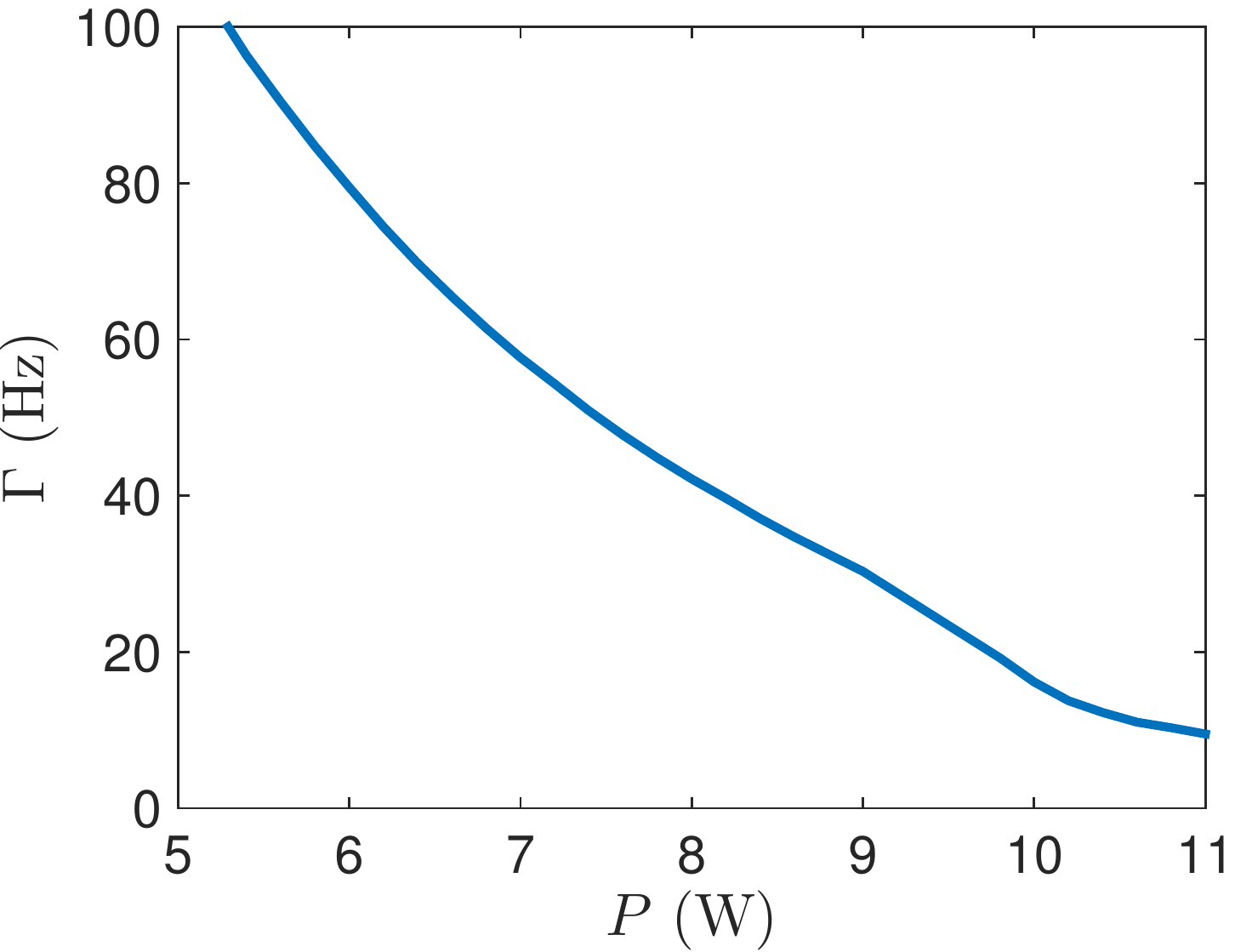}
\caption{Total losses after optimization of detunings as a function of total input power for trap depth of $10$ \textmu K.
}
\label{fig:optimized_losses}
\end{figure}

We formulate the loss optimization of an atomic trap as follows:
we estimate what is the minimal total loss of a 2D atomic trap when both total input power and the desired trap depth are fixed.
That is, given our waveguide geometry and desired trap depth, for each value of total input power we scan both blue and red detuned laser wavelengths in order to find optimal set of detunings for which the total loss is minimized.
Figure \ref{fig:optimized_losses} shows a characteristic optimization result, using the example of $V_\mathrm{trap} = 10$ \textmu K.
We emphasize here that optimized trap losses depend strongly on the propagating power, and can therefore be strongly reduced for laser written waveguides supporting large input powers.
In other words, for larger powers optimal detunings are further from atomic resonances, leading to smaller losses.
In Figure \ref{fig:optimized_losses}, we have limited the range of optimal detunings to start at least $5$nm away from atomic resonances in order to avoid possible molecular resonances for the red detuned laser \cite{Pichler:JCP:2004}.
At this point, we finally note that for total propagating power, $P_\mathrm{in} =8$ W,
the set of optimal detunings corresponds exactly to one used in Figure \ref{fig:two_mode_trap_opt},
i.e. $\delta \lambda_\mathrm{r} = 10.5$ nm, and $\delta \lambda_\mathrm{b} = 10$ nm.
Similar optimization analysis with different system parameters is straightforward, depending on chosen atomic trap properties (not shown here):
as expected, it will result in smaller optimized losses for smaller trap depths, and vice versa.

%%%
%%% Conclusion%%%

\section{Conclusion}
\label{sec:concl}

In conclusion, we have studied a promising interface for light-matter interactions based on laser written waveguides at the surface of fused silica. 
Our calculations demonstrate that trapping of atoms in evanescent light of waveguide modes is possible to achieve by superimposing two different laser frequencies in the waveguide.
Importantly, we have verified that atom trapping mechanism is successful in the regime where the blue detuned laser supports two propagating modes, whereas the red detuned laser operates in a single-mode regime. 
Unlike in the case of dielectric nanofibers \cite{Balykin+:PRA70:2004},
we have shown that the single-mode regime for both wavelengths is not expected to provide efficient trapping in laser written waveguide setup. 
We have focused on a particular example of Cs atoms with one wavelength blue detuned from atomic D2 resonance,
and the other wavelength red detuned from D1 resonance.
Numerical results are presented for reasonable experimental parameters: as an example we have investigated atomic traps with potential depth $\sim 10$ \textmu K, realized with several Watts of total propagating power.
Finally, we have provided with optimization scheme for system parameters in order to minimize trap losses due to photon scattering and surface losses.

%%%
%%% Acknowldgement%%%

\section*{Acknowledgements}

We are grateful for many fruitful discussions with I. Lesanovsky and L. Hackerm\"{u}ller.
We acknowledge financial support from the FET grant 295293 (QuILMI) of the 7th framework programme of the European Commission and the UK UK Engineering and Physical Science Research Councuil via the grant EP/M01326X/1.

%%%
%%% Reference%%%

\section*{References}

\bibliographystyle{iopart-num}
\bibliography{quilmi_bib}

\providecommand{\newblock}{}
\begin{thebibliography}{10}
\expandafter\ifx\csname url\endcsname\relax
  \def\url#1{{\tt #1}}\fi
\expandafter\ifx\csname urlprefix\endcsname\relax\def\urlprefix{URL }\fi
\providecommand{\eprint}[2][]{\url{#2}}

\bibitem{Kimble:Nature:2008}
Kimble H~J 2008 {\em Nature\/} {\bf 453} 1023--1030

\bibitem{Goban+:NatComm5:2014}
Goban A, Hung C~L, Yu S~P, Hood J~D, Muniz J~A, Lee J~H, Martin M~J, McClung
  A~C, Choi K~S, Chang D~E, Painter O and Kimble H~J 2014 {\em Nature
  Communications\/} {\bf 5}

\bibitem{Pichler+:PRA91:2015}
Pichler H, Ramos T, Daley A~J and Zoller P 2015 {\em Physical Review A\/} {\bf
  91} 042116--19

\bibitem{Balykin+:PRA70:2004}
Balykin V~I, Hakuta K, Le~Kien F, Liang J~Q and Morinaga M 2004 {\em Physical
  Review A\/} {\bf 70} 011401--4

\bibitem{Vetsch+:PRL104:2010}
Vetsch E, Reitz D, Sagu{\'e} G, Schmidt R, Dawkins S~T and Rauschenbeutel A
  2010 {\em Physical Review Letters\/} {\bf 104} 203603

\bibitem{SzameitNolte:JPB43:2010}
Szameit A and Nolte S 2010 {\em Journal of Physics B: Atomic, Molecular and
  Optical Physics\/} {\bf 43} 163001

\bibitem{Lederer+:PhysRep463:2008}
Lederer F, Stegeman G~I, Christodoulides D~N, Assanto G, Segev M and Silberberg
  Y 2008 {\em Physics Reports\/} {\bf 463} 1--126

\bibitem{Longhi:LPR3:2009}
Longhi S 2009 {\em Laser {\&} Photonics Reviews\/} {\bf 3} 243--261

\bibitem{Szameit:PRL:2007}
Szameit A, Kartashov Y~V, Dreisow F, Pertsch T, Nolte S, T{\"u}nnermann A and
  Torner L 2007 {\em Physical Review Letters\/} {\bf 98} 173903

\bibitem{Rechtsman+:NatPhot7:2012}
Rechtsman M~C, Zeuner J~M, T{\"u}nnermann A, Nolte S, Segev M and Szameit A
  2012 {\em Nature Photonics\/} {\bf 7} 153--158

\bibitem{Rechtsman+:Nature496:2013}
Rechtsman M~C, Zeuner J~M, Plotnik Y, Lumer Y, Podolsky D, Dreisow F, Nolte S,
  Segev M and Szameit A 2013 {\em Nature\/} {\bf 496} 196--200

\bibitem{Graefe+:NatPhot8:2014}
Gr{\"a}fe M, Heilmann R, Perez-Leija A, Keil R, Dreisow F, Heinrich M,
  Moya-Cessa H, Nolte S, Christodoulides D~N and Szameit A 2014 {\em Nat.
  Phot.\/} {\bf 8} 791--795

\bibitem{Maselli+:OE17:2009}
Maselli V, Grenier J~R, Ho S and Herman P~R 2009 {\em Optics Express\/} {\bf
  17} 11719--11729

\bibitem{LeKien+:PRA70:2004}
Le~Kien F, Balykin V~I and Hakuta K 2004 {\em Physical Review A\/} {\bf 70}
  063403--9

\bibitem{Lacroute+:NJP14:2012}
Lacro{\^u}te C, Choi K~S, Goban A, Alton D~J, Ding D, Stern N~P and Kimble H~J
  2012 {\em New Journal of Physics\/} {\bf 14} 023056

\bibitem{Goban+:PRL109:2012}
Goban A, Choi K~S, Alton D~J, Ding D, Lacro{\^u}te C, Pototschnig M, Thiele T,
  Stern N~P and Kimble H~J 2012 {\em Physical Review Letters\/} {\bf 109}
  033603

\bibitem{Alton+:NatPhys7:2010}
Alton D~J, Stern N~P, Aoki T, Lee H, Ostby E, Vahala K~J and Kimble H~J 2010
  {\em Nature Physics\/} {\bf 7} 159--165

\bibitem{Tiecke+:Nature508:2014}
Tiecke T~G, Thompson J~D, de~Leon N~P, Liu L~R, Vuleti{\'c} V and Lukin M~D
  2014 {\em Nature\/} {\bf 508} 241--244

\bibitem{Jukic+:SPIE9379:2015}
Juki{\'c} D, Pohl T and G{\"o}tte J~B 2015 {Evanescent fields of laser written
  waveguides} {\em SPIE OPTO\/} ed Galvez E~J, Gl{\"u}ckstad J and Andrews D~L
  (SPIE) pp 93790R--93790R--9

\bibitem{Steck:Cs:2010}
Steck D~A 2012 {Cesium D Line Data} Tech. rep. University of Oregon

\bibitem{Fallahkhair+:JLT26:2008}
Fallahkhair A~B, Li K~S and Murphy T~E 2008 {\em Journal of Lightwave
  Technology\/} {\bf 26} 1423--1431

\bibitem{Aiello+:PRL103:2009}
Aiello A, Lindlein N, Marquardt C and Leuchs G 2009 {\em Physical Review
  Letters\/} {\bf 103} 100401

\bibitem{DennisGoette:JOpt15:2013}
Dennis M~R and G{\"o}tte J~B 2013 {\em Journal of Optics\/} {\bf 15} 014015

\bibitem{Bliokh+:NatComm5:2014}
Bliokh K~Y, Bekshaev A~Y and Nori F 2014 {\em Nature Communications\/} {\bf 5}
  3300

\bibitem{Grimm+:AdvAMOPhys42:2000}
Grimm R, Weidemuller M and Ovchinnikov Y~B 2000 {\em Advances in Atomic
  Molecular, and Optical Physics, Vol. 42\/} {\bf 42} 95--170

\bibitem{LeKien+:EPJD67:2013}
Le~Kien F, Schneeweiss P and Rauschenbeutel A 2013 {\em The European Physical
  Journal D\/} {\bf 67} 92--16

\bibitem{Stern+:NJP13:2011}
Stern N~P, Alton D~J and Kimble H~J 2011 {\em New Journal of Physics\/} {\bf
  13} 085004

\bibitem{Pichler:JCP:2004}
Pichler M, Chen H and Stwalley W~C 2004 {\em J. Chem. Phys.\/} {\bf 121}

\end{thebibliography}

\end{document}